\documentclass[12pt,a4paper]{article}
\usepackage{amsmath}
\usepackage{amssymb}
\usepackage{enumerate}
\usepackage{amsfonts}
\usepackage[latin1]{inputenc}
\begin{document}

%
\newcounter{saveeqn}
\newcommand{\alpheqn}{\setcounter{saveeqn}{\value{equation}}%
\setcounter{equation}{0}%
\addtocounter{saveeqn}{1}
\renewcommand{\theequation}{\mbox{\arabic{saveeqn}\alph{equation}}}}
\newcommand{\reseteqn}{\setcounter{equation}{\value{saveeqn}}%
\renewcommand{\theequation}{\arabic{equation}}}

\begin{center}
{\bf Hamiltonian, Path Integral and BRST Formulations of Large N Scalar \\ $QCD_{2}$ on the Light-Front and Spontaneous Symmetry Breaking}
\\[14mm]

Usha Kulshreshtha$^{[{\rm a,b}]}$\footnote{Corresponding Author},~ Daya Shankar Kulshreshtha$^{[{\rm a,c}]}$,\\
and James P. Vary$^{[{\rm a}]}$ \footnote{Email Addresses: ushakulsh@gmail.com (U. Kulshreshtha), dskulsh@gmail.com (D. S. Kulshreshtha), jvary@iastate.edu (J. P. Vary)}

\end{center}

\begin{center} 

a Department of Physics and Astronomy,

  Iowa State University, Ames, Iowa 50011, USA \\

b Department of Physics, Kirori Mal College, 

  University of Delhi, Delhi-110007, India. \\

c Department of Physics and Astrophysics,

  University of Delhi, Delhi-110007, India \\	
  
\end{center} 

\vspace{2cm}

\begin{abstract}

Recently Grinstein, Jora, and Polosa have studied a theory of large-$N$ scalar quantum chromodynamics in  one-space one-time dimension. This theory admits a Bethe-Salpeter equation describing the discrete spectrum of  quark-antiquark bound states. They consider gauge fields in the adjoint representation of $SU(N)$  and scalar fields in the fundamental representation. The theory is asymptotically free and linearly confining. The theory could possibly provide a good field theoretic framework for the description of a large class of diquark-antidiquark (tetra-quark) states. Recently we have studied the light-front quantization of this theory without a Higgs potential.  In the present work, we study the light-front Hamiltonian, path integral and BRST formulations of the theory in the presence of a Higgs potential. The light-front theory is seen to be gauge-invariant, possessing a set of first-class constraints. The explicit occurrence of spontaneous symmetry breaking in the theory is shown in unitary gauge as well as in the light-front 't Hooft gauge.  
\end{abstract}

\newpage

\section{Introduction}

Study of multi-quark states in quantum chromodynamics (QCD) has been a subject of wide interest  \cite{1}-\cite{27}. Their interpretation remains a challenging task, and a number of phenomenological models \cite{1}-\cite{26} have been proposed to understand the various experimental observations.  Some of the notable heavier states \cite{9}-\cite{19} which do not fit into the standard classification of mesons (quark-antiquark ($q \bar q$) states) and baryons (three-quark states) \cite{1}-\cite{3} are the exotic charmonium-like $X, Y, Z$ resonances \cite{4}, \cite{9}-\cite{19}. Even some relatively lighter states \cite{20}-\cite{26} do not find proper interpretation within the standard classification of mesons and baryons \cite{20}-\cite{26}. 

Various possibilities for understanding hadron structure beyond the usual mesons and baryons \cite{3,4} have been considered in the literature \cite{1}-\cite{26}. Some \textit{exotic} states find a natural interpretation in terms of the four-quark or tetra-quark ($q\bar q q \bar q$) states \cite{3}, \cite{9}-\cite{26}. By now it is widely perceived that not only heavy states such as the $X ,Y, Z$ states have an \textit{exotic} structure  as tetra-quark states or diquark$(Q)$-antidiquark$(\bar{Q})$ states \cite{3}, \cite{9}-\cite{19}, but even some light scalar mesons could also be identified  as diquark-antidiquark ($Q\bar{Q}$) or tetra-quark systems \cite{20}-\cite{25}. 

In the first approximation, even the nonet formed by $f_{0}(980)$, $a_{0}(980)$, $\kappa(900)$, $\sigma(500)$ is interpreted as the lowest $Q\bar{Q}$ multiplet \cite{20}-\cite{25}, and the decuplet of scalar mesons with masses above 1 Gev, formed by $f_{0}(1370)$, $f_{0}(1500)$, $f_{0}(1710)$, $a_{0}(1450)$, $K_{0}(1430)$, is interpreted as the lowest $q\bar{q}$ scalar multiplet (cf. Refs.\cite{20}-\cite{25}). 

The multi-quark hadron states can be extremely broad \cite{9}-\cite{25}, and thus they could escape experimental identification. In this context the diquark-antidiquark structures \cite{3}, \cite{9}-\cite{25} have been suggested to explain several decay patterns of light scalar mesons \cite{20}-\cite{26}, heavy-light diquarks have also been introduced to study the X, Y, Z spectroscopy \cite{9}-\cite{19}.

Further, 't Hooft, Isidori, Maiani, Polosa and Riquer \cite{24} and others \cite{20}-\cite{25} have shown how one could explain the decays of the light scalar mesons by assuming a dominant diquark-antidiquark structure for the lightest scalar mesons \cite{20}-\cite{25}, where the diquark is being taken to be a spin zero anti-triplet color state \cite{20}-\cite{25}.  Grinstein, Jora and Polosa \cite{25} have studied a model of large-$N$ \textit{scalar} QCD \cite{20}-\cite{25} in  one-space and one-time dimension. Their model admits \cite{25} a Bethe-Salpeter equation describing the discrete spectrum of $q\bar{q}$ bound states \cite{20}-\cite{25}. 

The work of Grinstein et al. \cite{25} is seen to further support this hypothesis. In the work of Grinstein et al. \cite{25}, the gauge fields have been considered \cite{25} in the adjoint representation of $SU(N)$ and the scalar fields in the fundamental representation. The theory is asymptotically free and linearly confining \cite{25}. Different aspects of this theory have been studied by several authors in various contexts \cite{20}-\cite{26}. 

Also, because there is no spin-statistics connection in one space and one-time dimension, the \textit{spinor} $QCD_{2}$ is structurally similar to the \textit{scalar} $QCD_{2}$ \cite{5}-\cite{8}. It is therefore enough to consider the \textit{scalar} $QCD_{2}$ for a study of several aspects of $QCD_{2}$ \cite{5}-\cite{8}. The large-N behavior of \textit{scalar} $QCD_{2}$ has been studied in details by 't Hooft and others \cite{4}-\cite{8}.

In view of the above, the  motivations for our present studies could be easily highlighted. In the first place, the work of 't Hooft, Isidori, Maiani, Polosa and Riquer \cite{24} and others \cite{20}-\cite{25} has clearly shown as to how one could achieve a satisfactory explanation of light scalar meson decays by assuming a dominant diquark-antidiquark structure for the lightest scalar mesons \cite{20}-\cite{25} (where the diquark is being taken to be a spin zero anti-triplet color state). In this work, a coherent picture of scalar mesons as a mixture of tetra-quark states (dominating the lightest mesons) and heavy quark-antiquark states (dominating the heavier mesons) emerges \cite{24}. 

The studies of Grinstein, Jora and Polosa \cite{25} on the large-$N$ \textit{scalar} $QCD_{2}$ \cite{20}-\cite{25}, further support the hypothesis of 't Hooft, Isidori, Maiani, Polosa and Riquer \cite{24} and others \cite{20}-\cite{25}, about the assumption of a dominant diquark-antidiquark structure for the scalar mesons. The work of Grinstein, Jora and Polosa is based on the assumption that scalar $QCD_{2}$ with a large number of colors could be used to compute the mass spectrum as well as to estimate the mass of the first radial excitation of the lowest diquark-antidiquark scalar meson. They have applied a numerical procedure to solve the Bethe-Salpeter equations and compute the bound state discrete spectrum of this confining theory \cite{25}. They have even obtained the possible masses of the spinor and scalar quarks by imposing that the ratio of the ground state eigenvalues of the spinor and scalar Bethe-Salpeter equations respectively, is equal to the ratio of the physical masses $m_{\pi}/m_{\sigma}$ (cf. Ref. \cite{25}, for further details). They have even extended their discussion to the case of spin-one diquarks. 

The above studies of Grinstein, Jora and Polosa \cite{25},
based on the scalar QCD with a large number of colors \textit{in one-space one-time dimension}, clearly point towards some definite possibilities of gaining some insight, at least at the qualitative level, about the  physical tetra-quark states \textit{in three-space one-time dimension}. In addition to this, it may also be possible to study this theory, 
\textit{in three-space one-time dimension}, at a somewhat later point of time.

In view of the above, it seems reasonable to pursue these studies further. In fact, in a recent paper \cite{26}, we have studied the light-front (LF) quantization (LFQ) \cite{28}-\cite{36} of this theory (with a mass term for the complex \textit{scalar} (diquark) field but without the Higgs potential) on the hyperplanes defined by the equal light-cone time $\tau = x^{+} = (x^0 + x^1)/\sqrt{2} = $ constant \cite{30}-\cite{36}, using the Hamiltonian \cite{28} and path integral \cite{29}-\cite{31} formulations. 

In the present work, we study the LF Hamiltonian, path integral, and Becchi-Rouet-Stora and Tyutin (BRST) \cite{37}-\cite{39} formulations of this theory in the presence of a Higgs potential \cite{32}-\cite{36} under appropriate light-cone gauge-fixing conditions. The LF theory is seen to be gauge-invariant (GI), possessing a set of first-class constraints. We absorb the mass term for the complex scalar (diquark) field $~\phi~$ in the definition of our Higgs potential \cite{25}, \cite{26}, and then we study the action of the theory. 

One of the important motivations for introducing the Higgs potential is to study the aspects related to the spontaneous symmetry breaking (SSB)\cite{32}-\cite{34} in the theory. 
Another important motivation for introducing the Higgs potential in the theory is related to our long-term goal related to the study of this theory using the discrete light-cone (LC) quantization (DLCQ) along with the coherent state formalism  \cite{40}-\cite{48}, where we wish not only to study the aspects of the spontaneous symmetry breaking (SSB) but we also wish to make contact with the experimentally observational aspects of this theory using the LF Hamiltonian approach to study the two- and three- body relativistic bound state problems \cite{25}, \cite{40}-\cite{48}. This work therefore constitutes a part of our bigger project which  involves a study of some aspects related to the spontaneous symmetry breaking as well as to a study of its DLCQ using the coherent state formalism \cite{40}-\cite{48}, in the LF Hamiltonian approach to study the two- and three-body relativistic bound state problems \cite{40}-\cite{48}. 

In this sense one could think that the theory under our present consideration could perhaps provide a good basic field theoretic framework for a study of a large class of diquark-antidiquark or the tetra-quark states \cite{2, 3}, \cite{9}-\cite{26} which have been investigated in various experiments. These are some of the motivations that necessitate our present studies. 

Now, because the theory is GI, we also study its BRST quantization \cite{37}-\cite{39} under appropriate BRST light-cone gauge-fixing. Usually in the Hamiltonian and the path integral quantization of a theory  the gauge-invariance of the theory  gets broken because the procedure of gauge-fixing converts the set of first-class constraints of the theory into a set of second-class constraints. A possible way to achieve the quantization of a GI theory, such that the gauge-invariance of the theory is maintained even under gauge-fixing is to use a generalized procedure called the BRST quantization \cite{37}-\cite{39}, where the extended gauge symmetry, called the BRST symmetry, is maintained even under gauge-fixing.

In the next section, we briefly recap some basics of the instant-form (IF) quantization (IFQ) of this theory in the presence of a Higgs potential. Its LFQ in the presence of a Higgs potential is then considered in Sec. 3, using the Hamiltonian and path integral formulations. The light-front BRST formulation of the theory is studied in Sec. 4, under the appropriate BRST light-cone gauge-fixing. Finally, the summary and discussion are given in Sec. 5. 

\section{Some Basics of the Theory} 

In this section we recap some of the basics of this theory of large-$N$ \textit{scalar} $QCD_{2}$ \textit{in the presence} of a Higgs potential, studied earlier by Grinstein, Jora and Polosa \textit{without} a Higgs potential (but with a mass term for the complex \textit{scalar} (diquark) field $\phi ~$) \cite{25} (the mass term for the complex \textit{scalar} (diquark) field $~\phi ~$ is absorved in the definition of our Higgs potential \cite{25, 26}). The theory of large-$N$ \textit{scalar} $QCD_{2}$ that we propose to study is defined by the action \cite{25}:
\begin{eqnarray}
S &=& \int_{ }^{} {\cal L} (\phi, \phi^{\dagger}, A^{\mu}) 
d^{2} x \notag \\
{\cal L} &=& \biggl[-\frac{1}{4} F_{\mu\nu} F^{\mu\nu}  + \partial_{\mu}\phi^{\dagger} \partial^{\mu}\phi + [i \rho (\phi A^{\mu} \partial_{\mu}\phi^{\dagger} - \phi^{\dagger}A_{\mu} \partial^{\mu} \phi) + \rho^{2}\phi^{\dagger}\phi A_{\mu} 
A^{\mu}] - V(|\phi|^{2}) \biggr] \notag \\
|\phi|^{2} &=& \phi^{\dagger}\phi
~,~ V(|\phi|^{2}) = \biggl[\mu^{2} (\phi^{\dagger}\phi) + \frac{\lambda}{6} (\phi^{\dagger}\phi)^{2} \biggr] 
~,~ \phi_{0}\ne 0 ~,~ (-\mu^{2} > 0 ~,~ \lambda > 0 ) \notag \\
g^{2} &:=& (4 \pi \alpha_{s}) ~~~~,~~~~ \rho = \frac{g}{\sqrt{N}} \notag \\ 
g^{\mu\nu} &=& g_{\mu\nu} :=  \left(\begin{array}{cc}  1 & 0 \\ 0 & -1 \end{array} \right) ,~  \mu , ~\nu = 0,~1 ~~~~ (IF) \notag \\  
g^{\mu\nu} &=& g_{\mu\nu} :=  \left(\begin{array}{cc} 0 & 1 \\ 1 & 0 \end{array} \right) , ~~ \mu ,~\nu = +, ~ - ~~~~ (FF) 
\label{1}
\end{eqnarray}
Here $\alpha_{s}$ is the QCD coupling constant.
The covariant derivative in our considerations is defined as:
\begin{equation}
D_{\mu} = (\partial_{\mu} + i \rho A_{\mu}^{a} T^{a}) 
= (\partial_{\mu} + i \rho A_{\mu}) 
\label{2}
\end{equation}
where $A_{\mu} (\equiv A_{\mu}^{a} T^{a})$ are the gluon gauge fields and $~ T^{a} ~$ are the generators of Lie algebra corresponding to the group $SU(N_{c})$ obeying the commutation relations: 
\begin{eqnarray}
[ T^{a} ~,~ T^{b} ] = i f^{abc} T^{c}  ~~~;~~~
a,~ b ,~ c  = 1,2,......, (N_{c}^{2} - 1) 
\label{3}
\end{eqnarray}
with $~ N_{c} = 2 ~$ for $~ SU(2) ~$ and $~ N_{c} = 3 ~$ for 
$~ SU(3) ~$. The structure constants $f^{abc}$ are antisymmetric in all indices. The gluon gauge field strength $F_{\mu\nu}^{a}$ is defined as: 
\begin{eqnarray}
F_{\mu\nu}^{a} &=& [(\partial_{\mu} A_{\nu}^{a} - \partial_{\nu} A_{\mu}^{a}) + \rho (A_{\mu} \times A_{\nu})^{a}]\notag \\
&=& [(\partial_{\mu} A_{\nu}^{a} - \partial_{\nu} A_{\mu}^{a}) + \rho f^{abc} A_{\mu}^{b} A_{\nu}^{c}]
\label{4}
\end{eqnarray} 
Here $~ (A_{\mu} \times A_{\nu})^{a} = f^{abc} A_{\mu}^{b} A_{\nu}^{c} ~$ defines the cross product for any two ``isotopic'' vectors: $~ A_{\mu}^{a} ~$ and $~ A_{\nu}^{a} ~$ \cite{8}. 

Further, the \textit{scalar} fields $\phi$ and $\phi^{\dagger}$ transform as the $N$ and $\bar{N}$ representations of the $U(N)$ color group respectively \cite{2}. Also, following the work of Grinstein, Jora and Polosa \cite{25}, we ignore all gluon self-coupling terms that arise from our chosen Lagrangian.

In the Lagrangian density of our theory (defined by Eq.(1)), the first term represents the kinetic energy of the gluon field, the second term represents the kinetic energy term for the \textit{scalar} (diquark) field, the third term represents the interaction term for the  \textit{scalar} (diquark) field with the gluon field, and the last term represents the Higgs potential which is kept rather general, without making any specific choice for the parameters $\mu^{2}$ and $\lambda$. However, they are chosen such that the potential remains a double well potential with the vacuum expectation value $\phi_{0} = <0|\phi(x)|0> \ne 0$, so as to allow for the spontaneous symmetry breaking in the theory. Also, the mass term for the \textit{scalar} (diquark) field has been absorbed in the definition of the Higgs potential. The values: $\mu^{2} = m^{2}$ and $\lambda = 0$ reproduce the theory of Grinstein, Jora and Polosa \cite{25}. 

The Euler-Lagrange equations of motion of the theory 
(with $\mu~,~ \nu = 0,~1 $ for IFQ and $\mu~,~ \nu = +,~- $ for LFQ) are obtained as:
\begin{eqnarray}
[\partial_{\mu} F^{\mu\nu} + i \rho (\phi \partial^{\nu} \phi^{\dagger} -
\phi^{\dagger} \partial^{\nu} \phi)
+ 2 \rho^{2} \phi^{\dagger} \phi A^{\nu} ] &=& 0 
\notag \\
\biggl[-\mu^{2} \phi^{\dagger} - \frac{\lambda}{3} (\phi^{\dagger}\phi) \phi^{\dagger}
+ \rho^{2} \phi^{\dagger} A_{\mu} A^{\mu} + i \rho A_{\mu} \partial^{\mu} \phi^{\dagger} 
+ i \rho \partial_{\mu} (\phi^{\dagger} A^{\mu}) 
 - \partial_{\mu} \partial^{\mu} \phi^{\dagger} 
\biggr] &=& 0 
\notag \\
\biggl[-\mu^{2} \phi - \frac{\lambda}{3} (\phi^{\dagger}\phi) \phi + \rho^{2} \phi A_{\mu} A^{\mu} - i \rho A_{\mu} \partial^{\mu} \phi 
- i \rho \partial_{\mu} (\phi A^{\mu}) 
- \partial_{\mu} \partial^{\mu} \phi \biggr] &=& 0 
\label{5}
\end{eqnarray}

\section{Instant-Form Quantization} 

We now consider the instant-form (IF) quantization (IFQ) of the theory. The action of the above theory in the IF of dynamics (with $A_{0} \equiv A^{a}_{0} T^{a} ~,~ A_{1} \equiv A^{a}_{1} T^{a}$) reads \cite{25}: 
\begin{eqnarray}
S &=& \int_{ }^{}  {\cal L}~dtdx  \notag \\
{\cal L} &=& \biggl[\frac{1}{2} (\partial_{0}A_{1}  - \partial_{1} A_{0} )^{2} 
+ (\partial_{0}\phi^{\dagger} \partial_{0}\phi - \partial_{1}\phi^{\dagger} \partial_{1}\phi)
+ \rho^{2} \phi^{\dagger} \phi (A_{0}^2  -  A_{1}^2 ) 
\notag \\
&& ~ + i \rho (\phi A_{0}  \partial_{0}\phi^{\dagger} - \phi A_{1}  \partial_{1}\phi^{\dagger} - 
\phi^{\dagger} A_{0}  \partial_{0}\phi + \phi^{\dagger} A_{1}  \partial_{1}\phi) - \mu^{2}(\phi^{\dagger}\phi) - \frac{\lambda}{6} (\phi^{\dagger}\phi)^{2} \biggr] 
\label{6}
\end{eqnarray}
Here $t = x^{0} = x_{0}$ and $x = x^{1} = - x_{1}$. Canonical momenta obtained from the above action  are:
\begin{eqnarray}
\pi &:=& \frac{\partial{\cal L}}{\partial(\partial_{0}\phi)}
= (\partial_{0}\phi^{\dagger} - i \rho A_{0}  \phi^{\dagger}) ~,~
\pi^{\dagger} := \frac{\partial{\cal L}}{\partial(\partial_{0}\phi^{\dagger})}
= (\partial_{0}\phi + i \rho A_{0} \phi) \notag \\
\Pi^{0} &:=& \frac{\partial{\cal L}}{\partial(\partial_{0}A_{0} )} = 0 ~,~
E (= \Pi^{1} ) := \frac{\partial{\cal L}}
{\partial(\partial_{0}A_{1} )}
= (\partial_{0} A_{1}  - \partial_{1} A_{0} ) 
\label{7}
\end{eqnarray}
Here $\pi , \pi^{\dagger}$ , $\Pi^{0} (\equiv \Pi^{0 a} T^{a})$ and $E := \Pi^{1} (\equiv \Pi^{1 a} T^{a})$ are the momenta canonically conjugate respectively to $\phi, \phi^{\dagger}, A_{0}$ and $A_{1}$. The above equations however, imply that the theory possesses only one primary constraint:
\begin{equation}
\chi_{1} = \Pi^{0} \approx 0 
\label{8}
\end{equation}
The symbol $ \approx $  here denotes a weak equality in the sense of Dirac \cite {28}, and it implies that the  constraints hold as a strong equality only on the reduced hyper surface of the constraints and not in the rest of the  phase space of the classical theory (and similarly one can consider it as a weak operator equality for the corresponding quantum theory). 

The canonical Hamiltonian density corresponding to ${\cal L}$ is:
\begin{eqnarray}
{\cal H}_{c} &:=& \biggl[\pi \partial_{0}\phi  + \pi^{\dagger} \partial_{0}\phi^{\dagger} + \Pi^{0} \partial_{0}A_{0}  + E  \partial_{0}A_{1} 
- {\cal L} \biggr]  
\notag  \\
&=& \biggl[\frac{1}{2}(E)^{2}  - A_{0} \partial_{1}E + \pi^{\dagger}\pi+ \partial_{1} \phi^{\dagger} \partial_{1}\phi + \rho^{2} A_{1}^{2} \phi^{\dagger}\phi \nonumber \\
&& ~~~~~~ - i \rho A_{0}(\phi\pi - \phi^{\dagger}\pi^{\dagger}) 
 - i \rho A_{1} (\phi^{\dagger} \partial_{1}\phi - \phi \partial_{1}\phi^{\dagger}) 
+ \mu^{2}(\phi^{\dagger}\phi) + \frac{\lambda}{6} 
(\phi^{\dagger}\phi)^{2} \biggr]   
\label{9}
\end{eqnarray}
After including the primary constraint $\chi_{1}$ in the canonical Hamiltonian density ${\cal H}_{c}$ with the help of the Lagrange multiplier field  $ u $, the total Hamiltonian density ${\cal H}_{T}$ could be written as :
\begin{eqnarray}
{\cal H}_{T}&=& \biggl[ \Pi^{0} u + \frac{1}{2}(E)^{2}  - A_{0}\partial_{1}E  
+ \pi^{\dagger}\pi+ \partial_{1} \phi^{\dagger} \partial_{1}\phi + \rho^{2} A_{1}^{2} \phi^{\dagger}\phi \notag \\
&& ~~~~~~ - i \rho A_{0}(\phi\pi - \phi^{\dagger}\pi^{\dagger}) 
 - i \rho A_{1} (\phi^{\dagger} \partial_{1}\phi - \phi \partial_{1}\phi^{\dagger}) 
+ \mu^{2}(\phi^{\dagger}\phi) + \frac{\lambda}{6} 
(\phi^{\dagger}\phi)^{2} \biggr]  
\label{10}
\end{eqnarray}
Hamilton's equations of motion of the theory that preserve the constraints of the theory in the course of time could be obtained from the total Hamiltonian: 
$H_{T}= \int_{ }^{} {\cal H}_{T}dx^{1}$ (and are omitted here for the sake of brevity). Demanding that the primary constraint $\chi_{1}$ be preserved in the course of time, one obtains the secondary Gauss-law constraint of the theory as:
\begin{equation}
\chi_{2} = [\partial_{1} E + i \rho (\phi\pi - \phi^{\dagger} \pi^{\dagger})] \approx 0 
\label{11}
\end{equation}
The preservation of $\chi_{2}$ for all times gives rise to one further constraint:
\begin{equation}
\chi_{3} = [2 \rho^{2} A_{0} \pi^{\dagger} \phi^{\dagger} + i \rho A_{1}(\phi \partial_{1} \phi^{\dagger} 
+ \phi^{\dagger} \partial_{1}\phi)] \approx 0 
\label{12}
\end{equation}
The theory is thus seen to possess only three constraints $\chi_{i}$ (with i = 1,2,3). The matrix $R_{\alpha\beta}$ of the Poisson brackets among the set of constraints $\chi_{i}$ with $( i = 1,2,3)$ is seen to be singular (the other details of the matrix $R_{\alpha\beta}$ are omitted here for the sake of brevity). This implies that the set of constraints $\chi_{i}$ is first-class and that the theory under consideration is gauge-invariant (GI). Consequently the theory is seen to possess the local vector gauge symmetry defined by the local vector gauge transformations:
\begin{equation}
\delta \phi = i \rho \beta \phi ~,~ \delta \phi^{\dagger} = - i \rho \beta \phi^{\dagger} ~,~ 
\delta  A_{0} =  \partial_{0} \beta ~,~ \delta A_{1} = \partial_{1} \beta 
\label{13}
\end{equation}
where $\beta \equiv \beta (x_{0},x_{1})$ is an arbitrary real function of its arguments. This theory could now be quantized under some appropriate gauge-fixing conditions, e.g., under the time-axial or temporal gauge: $ A_{0} \approx 0 $. The details of this IFQ are however, outside the scope of the present work (what actually happens is that one of the  matrix elements of the matrix $R_{\alpha\beta}$ involves a linear combination of a Dirac distribution function $\delta (x^1 - y^1)$ and its first derivative and finding its inverse is a rather non-trivial task). We now proceed with the LFQ of this theory in the next section.

\section{Light-Front Hamiltonian and Path Integral \\Quantization} 

In this section we study the LF Hamiltonian and path integral formulations \cite{25}-\cite{31} of the above theory \cite{25} under appropriate LC gauge-fixing.  The action for the scalar theory in LF coordinates 
$ x^{\pm} := (x^{0} \pm x^{1})/ \sqrt{2} ~~$ 
(with $A^{+} \equiv A^{+ a} T^{a} ~,~ A^{-} \equiv A^{- a}T^{a}$) reads: 
\begin{eqnarray}
S &=& \int_{ }^{}  {\cal L} ~ dx^{+}dx^{-}  \notag \\
{\cal L} &=& \biggl[\frac{1}{2} (\partial_{+}A^{+}  - \partial_{-} A^{-})^{2} 
+ (\partial_{+}\phi^{\dagger}\partial_{-}\phi + \partial_{-}\phi^{\dagger}\partial_{+}\phi) 
- \mu^{2} (\phi^{\dagger}\phi) - \frac{\lambda}{6} (\phi^{\dagger}\phi)^{2} \notag \\
&& ~~~~ + i \rho A^{+} (\phi \partial_{+}\phi^{\dagger} - \phi^{\dagger} \partial_{+}\phi) + i \rho A^{-} (\phi\partial_{-}\phi^{\dagger} - \phi^{\dagger} \partial_{-}\phi) + 2 \rho^{2} \phi^{\dagger}\phi A^{+}  A^{-} 
\biggr] 
\label{14}
\end{eqnarray}
In the work of Ref.\cite{25}, the authors have studied the above action, after implementing the  gauge-fixing condition (GFC) $A^{+} \approx 0$ ``strongly'' in the above action.  In contrast to this, we propose to study the theory defined by the above action, following the standard Dirac quantization procedure \cite{28} and we do not fix any gauge at this stage. We instead consider this gauge-fixing condition $(A^{+} \approx 0)$ as one of the gauge constraints \cite{28}-\cite{31} which becomes strongly equal to zero only on the reduced hyper surface of the constraints and remains non-zero in the rest of the phase space of the theory. 
 
It may be important to note here that one of the salient features of Dirac quantization procedure \cite{28} is that in this quantization the gauge-fixing conditions should be treated on par with other gauge-constraints of the theory which are only weakly equal to zero in the sense of Dirac\cite{28}, and they become strongly equal to zero only on the reduced hyper surface of the constraints of the theory and not in the rest of the phase space of the classical theory (in the corresponding quantum theory these weak equalities  become the weak operator equalities). 

Another thing to be noted here is that we have introduced the Higgs potential in our present work and we have absorbed the mass term for the scalar (diquark) field in the definition of our Higgs potential \cite{25,26}, \cite{32}-\cite{34}. We now proceed to study the LF Hamiltonian and path integral formulations of the theory defined by the above action. The LF Euler-Lagrange equations of motion of the theory are: 
\begin{eqnarray}
\biggl[(\partial_{+}\partial_{-}A^{-}  - \partial_{+}\partial_{+}A^{+} )  
+ i \rho (\phi \partial_{+}\phi^{\dagger} -
\phi^{\dagger} \partial_{+} \phi ) +  2 \rho^{2} \phi^{\dagger}\phi A^{-} \biggr] &=& 0 
\notag \\
\biggl[(\partial_{+}\partial_{-}A^{+} - \partial_{-} \partial_{-} A^{-})   + i \rho (\phi \partial_{-}\phi^{\dagger} - \phi^{\dagger} \partial_{-} \phi) + 2 \rho^{2} \phi^{\dagger}\phi A^{+} \biggr] &=& 0 
\notag \\
\biggl[- \mu^{2} \phi^{\dagger} - \frac{\lambda}{3} 
\phi^{\dagger}\phi \phi^{\dagger}
 - 2\partial_{+} \partial_{-} \phi^{\dagger} + 2 i \rho (A^{+} \partial_{+}\phi^{\dagger} + A^{-} \partial_{-} \phi^{\dagger}) ~~~~ ~~~~ ~~ \notag \\
 + i\rho \phi^{\dagger}(\partial_{+} A^{+} 
- \partial_{-} A^{-}) + 2 \rho^{2} \phi^{\dagger} A^{+}  A^{-} \biggr] &=& 0 
\notag \\ 
\biggl[- \mu^{2} \phi - \frac{\lambda}{3} \phi^{\dagger}\phi \phi - 2 \partial_{+} \partial_{-} \phi - 2 i \rho (A^{+} \partial_{+}\phi + A^{-}  \partial_{-} \phi) ~~~~ ~~~~ ~~ \notag \\
- i \rho \phi (\partial_{+} A^{+} - \partial_{-} A^{-}) + 2 \rho^{2} \phi A^{+}  A^{-} \biggr] &=& 0 
\label{15}
\end{eqnarray}
In the following, we consider the Hamiltonian formulation of the theory described by the above  action. The canonical momenta obtained from the above action  are:
\begin{eqnarray}
\pi &:=& \frac{\partial{\cal L}}{\partial(\partial_{+}\phi)}
= (\partial_{ -}\phi^{\dagger} - i \rho A^{+} \phi^{\dagger}) ~,~
\pi^{\dagger} := \frac{\partial{\cal L}}{\partial(\partial_{+}\phi^{\dagger})}
= (\partial_{-}\phi + i \rho A^{+}  \phi) \notag \\
\Pi^{+} &:=& \frac{\partial{\cal L}}{\partial(\partial_{+}A^{-})} = 0 ~,~
\Pi^{-} := \frac{\partial{\cal L}}{\partial(\partial_{+}A^{+})}
= (\partial_{+} A^{+} - \partial_{-} A^{-}) 
\label{16}
\end{eqnarray}
Here $\pi , \pi^{\dagger}$ , $\Pi^{+} (\equiv \Pi^{+ a} T^{a})$ and $\Pi^{-} (\equiv \Pi^{- a} T^{a})$  are the momenta canonically conjugate respectively to $\phi , \phi^{\dagger} ,  A^{-}$ and $A^{+}~.$ 

The above equations however, imply that the theory possesses three primary constraints:
\begin{equation}
\chi_{1} = \Pi^{+} \approx 0 ~~,~~
\chi_{2} = [\pi - \partial_{-}\phi^{\dagger} + i \rho 
A^{+} \phi^{\dagger}] \approx 0  
~~,~~ 
\chi_{3} = [\pi^{\dagger} - \partial_{-}\phi - i \rho 
A^{+} \phi ]\approx 0  
\label{17}
\end{equation}
The canonical Hamiltonian density corresponding to ${\cal L}$ is:
\begin{eqnarray}
{\cal H}_{c}&=&\biggl[\pi \partial_{+}\phi + \pi^{\dagger}\partial_{+}\phi^{\dagger}+ \Pi^{+}  \partial_{+}A^{-}  + \Pi^{-} \partial_{+}A^{+} - {\cal L}\biggr]  \notag \\
&=& \biggl[\frac{1}{2}(\Pi^{-})^{2} + \Pi^{-} (\partial_{-} A^{-}) + \mu^{2} (\phi^{\dagger}\phi) + \frac{\lambda}{6} (\phi^{\dagger}\phi)^{2} 
\notag \\
&& ~~~~~~~~ - i \rho  A^{-} (\phi \partial_{-} \phi^{\dagger} - \phi^{\dagger} \partial_{-} \phi) - 2 \rho^{2} \phi^{\dagger}\phi A^{+} A^{-} \biggr]
\label{18}
\end{eqnarray}
After including the primary constraints $\chi_{1} , \chi_{2}$ and $\chi_{3}$ in the canonical Hamiltonian density ${\cal H}_{c}$ with the help of the Lagrange multiplier fields  $ u , v $ and $w$,   the total Hamiltonian density ${\cal H}_{T}$ could be written as :
\begin{eqnarray}
{\cal H}_{T}&=&\bigg[(\Pi^{+}) u + (\pi - 
\partial_{-}\phi^{\dagger} + i \rho A^{+} \phi^{\dagger}) v + (\pi^{\dagger} - \partial_{-}\phi - i \rho A^{+} \phi) w + \mu^{2} (\phi^{\dagger}\phi) 
\notag \\
&& + \frac{\lambda}{6} (\phi^{\dagger}\phi)^{2} 
 + \frac{1}{2}(\Pi^{-} )^{2} + \Pi^{-} \partial_{-} A^{-}   - i \rho  A^{-} (\phi \partial_{-} \phi^{\dagger} - \phi^{\dagger} \partial_{-} \phi) - 2 \rho^{2} \phi^{\dagger}\phi A^{+} A^{-}  \biggr]~~
\label{19}
\end{eqnarray}
The Hamilton's equations of motion of the theory that preserve the constraints of the theory in the course of time could be obtained from the total Hamiltonian (and are omitted here for the sake of brevity): $ H_{T}= \int_{ }^{} {\cal H}_{T}dx^{-} $. Demanding that the primary constraint $\chi_{ 1 }$ be preserved in the course of time, one obtains the secondary Gauss-law constraint of the theory as:
\begin{equation}
\chi_{4} = [ \partial_{-} \Pi^{-} + i \rho (\phi \partial_{-} \phi^{\dagger} - \phi^{\dagger} \partial_{-} \phi) +  2 \rho^{2} \phi^{\dagger}\phi A^{+} ] 
\approx 0
\label{20}
\end{equation}
The preservation of $\chi_{2},  \chi_{ 3} $ and   $\chi_{ 4}$, for all times does not give rise to any further constraints. The theory is thus seen to possess only four  constraints $\chi_{ i}$ (with i = 1,2,3,4). The constraints $\chi_{2}, \chi_{3}$ and $\chi_{ 4}$ could however, be combined in to a single constraint: 
\begin{equation}
\psi = [\partial_{-} \Pi^{-} + i \rho (\phi \pi
- \phi^{\dagger} \pi^{\dagger})] \approx 0
\label{21}
\end{equation}
and with this modification, the new \textit{set} of constraints of the theory could 
be written as:
\begin{equation}
\Omega_{1} = \chi_{1} = \Pi^{+} \approx 0 ~,~ 
\Omega_{2} = \psi = [\partial_{-} \Pi^{-} + i \rho (\phi \pi - \phi^{\dagger} \pi^{\dagger})] \approx 0
\label{22}
\end{equation}
Further, the matrix of the Poisson brackets among the constraints $\Omega_{i} $ , with $ (i = 1,2) $ is seen to be a singular matrix implying that the set of constraints $\Omega_{i}$ is first-class and that the theory under consideration is gauge-invariant. Expressions for the components of the vector gauge current density of the theory are obtained as:
\begin{eqnarray}
j^{+}&=& [ - i \rho \beta \phi \partial_{-} \phi^{\dagger} + i \rho \beta \phi^{\dagger} \partial_{-} \phi 
- 2 \rho^{2} \beta A^{+} \phi^{\dagger} \phi
- \beta (\partial_{-} \partial_{+} A^{+} - \partial_{-} \partial_{-} A^{-}) ] \notag  \\
j^{-} &=& [- i\rho \beta \phi \partial_{+}\phi^{\dagger} + i\rho \beta \phi^{\dagger}\partial_{+}\phi 
- 2 \rho^{2} \beta A^{-} \phi^{\dagger} \phi 
+ \beta (\partial_{+} \partial_{+} A^{+} - \partial_{+} \partial_{-} A^{-} ) ]   
\label{23}
\end{eqnarray}
The divergence of the vector gauge current density of the theory could now be easily seen to vanish satisfying the continuity equation: $\partial_{\mu} j^{\mu} = 0$ , implying that the theory possesses at the classical level, a local vector-gauge symmetry. The action of the theory  is indeed seen to be invariant under the local vector gauge transformations:
\alpheqn
\begin{eqnarray}
\delta \phi &=& - i \rho \beta \phi ~,~ \delta \phi^{\dagger} = i \rho \beta \phi^{\dagger} ~,~ \delta 
A^{-}  
= \partial_{+} \beta ~,~ \delta A^{+}  = \partial_{-} 
\beta  \\
\delta \pi &=& [{\rho}^{2} \beta \phi^{\dagger} A^{+} 
+ i\rho \beta \partial_{-}\phi^{\dagger}] 
  ~,~ \delta \pi^{\dagger} = [{\rho}^{2} \beta \phi A^{+} - i \rho \beta \partial_{-}\phi] \\
\delta u &=& \delta v = \delta w = \delta \Pi^{+}  = \delta \Pi^{-}  = \delta \Pi_{u} = \delta \Pi_{v} = \delta \Pi_{w} = 0 
\label{24}
\end{eqnarray}
\reseteqn
where $\beta \equiv \beta (x^{+},x^{-})$ is an arbitrary real function of its arguments and $\Pi_{u}, \Pi_{v}$ and $\Pi_{w}$ are the momenta canonically conjugate to the Lagrange multiplier fields $u, v$ and $w$ respectively, which are treated here as dynamical fields.
Using the Euler-Lagrange equations of motion of the theory and the expressions for the components of the vector gauge current density of the theory, one could now easily show that:
\begin{equation}
j^{+} = \beta (1+ \rho) [\partial_{-} \partial_{-} 
A^{-} - \partial_{+} \partial_{-} A^{+} ]
\label{25}
\end{equation}
It may be important to point out here that Grinstein, Jora and Polosa \cite{25}, have obtained an equation 
(under the gauge $A^{+} = 0$)  analogous to the above equation connecting $\partial_{-} \partial_{-} A^{-}$ and $j^{+}$
(cf. Eq. (5) of Ref. \cite{25}) which has been shown \cite{25}, to admit a solution (in the absence of background fields) \cite{25}, which when substituted in to the Lagrangian density of the theory implies a linear potential between charges (for further details, we refer to the work of Ref.\cite{25}). 

In order to quantize the theory using Dirac's procedure we now convert the set of first-class constraints of the theory  $\eta_{i}$ into a set of second-class constraints, by imposing, arbitrarily, some additional constraints on the system called gauge-fixing conditions (GFC's) or the gauge-constraints \cite{28}-\cite{31}. For the present theory, we could choose, for example, the following set of GFC's: $\zeta_{1} = A^{+}  \approx 0 ~$, $~ \zeta_{2} = A^{-}  \approx 0$. Here the gauge $A^{+} \approx 0$ represents the LC time-axial or temporal gauge and the gauge $A^{-}  \approx 0$ represents the LC coulomb gauge and both of these gauges are physically important gauges. Corresponding to this gauge choice, the theory has  the following  set of constraints under which the quantization of the theory could for example be studied:
\alpheqn
\begin{eqnarray}
\xi_{1} &=& \Omega_{1} =  \chi_{1} = \Pi^{+} \approx 0  \\
\xi_{2} &=& \Omega_{2} = \psi = [\partial_{-} \Pi^{-} + i \rho (\phi \pi
- \phi^{\dagger} \pi^{\dagger})] \approx 0 \\
\xi_{3} &=& \zeta_{1} = A^{+}  \approx 0 \\
\xi_{4} &=& \zeta_{2} = A^{-} \approx 0 
\label{26}
\end{eqnarray}
\reseteqn
The matrix $R_{\alpha\beta} $ of the Poisson brackets among the set of constraints $\xi_{ i}$ with $( i = 1,2,3,4) $ is seen to be nonsingular with the determinant given by
\begin{eqnarray}
\biggl[||det(R_{\alpha\beta})|| \biggr]^\frac{1}{2} &=&  \biggl[
\partial_{-}\delta (x^{-} - y^{-}) ~~ \delta (x^{-} - y^{-}) \biggr]
\label{27}
\end{eqnarray}
The other details of the matrix $R_{\alpha\beta} $ are omitted here for the sake of brevity. Finally, following  the Dirac quantization procedure, the nonvanishing equal light-cone-time commutators of the theory, under the GFC's: $ A^{+} \approx 0 $ and  
$ A^{-} \approx 0 $ are obtained as:
\alpheqn
\begin{eqnarray}
[\phi(x^{+}, x^{-})~ ,~ \pi(x^{+}, y^{-})] &=& 
i ~ \delta(x^{-} - y^{-}) \\
\label{28a}
[\phi^{\dagger}(x^{+}, x^{-})~ ,~ \pi^{\dagger}(x^{+}, y^{-})] &=& i~\delta(x^{-} - y^{-}) \\
\label{28b}
[\phi(x^{+}, x^{-})~ ,~ \Pi^{-} (x^{+}, y^{-})] &=& 
 ~ \frac{1}{2} \rho \phi ~ \epsilon(x^{-} - y^{-}) \\
\label{28c}
[\phi^{\dagger} (x^{+}, x^{-})~ ,~ \Pi^{-} (x^{+}, y^{-})] &=& - ~ \frac{1}{2} \rho \phi^{\dagger} ~ \epsilon(x^{-} - y^{-}) \\
\label{28d}
[\pi(x^{+}, x^{-})~ ,~ \Pi^{-} (x^{+}, y^{-})] &=& 
\frac{1}{2} ~\rho ~ \pi~ \epsilon (x^{-} - y^{-}) \\
\label{28e}
[\pi^{\dagger}(x^{+}, x^{-})~ ,~ \Pi^{-} (x^{+}, y^{-})] 
&=& - ~\frac{1}{2} \rho ~ \pi^{\dagger} ~ \epsilon(x^{-} - y^{-}) \\
\label{28f}
[\Pi^{-} (x^{+}, x^{-})~ ,~ \phi(x^{+}, y^{-})] &=& 
 ~ \frac{1}{2} \rho \phi ~\epsilon(x^{-} - y^{-}) \\
\label{28g}
[\Pi^{-} (x^{+}, x^{-})~ ,~ \phi^{\dagger}(x^{+}, y^{-})] 
&=& - ~ \frac{1}{2} \rho \phi^{\dagger} ~\epsilon(x^{-} - y^{-}) \\
\label{28h}
[\Pi^{-} (x^{+}, x^{-})~ ,~ \pi(x^{+}, y^{-})] 
&=& - ~ \frac{1}{2} \rho \pi ~\epsilon(x^{-} - y^{-}) \\
\label{28i}
[\Pi^{-} (x^{+}, x^{-})~ ,~ \pi^{\dagger}(x^{+}, y^{-})] 
&=& ~ \frac{1}{2} \rho \pi^{\dagger} ~\epsilon(x^{-} - y^{-})
\label{28j}
\end{eqnarray}
\reseteqn
The first-order Lagrangian density ${\cal L}_{I0}$ of the theory is:
\begin{eqnarray}
{\cal L}_{I0} & := & \biggl [ \pi (\partial _{+}\phi) + \pi^{\dagger}(\partial _{+}\phi^{\dagger})  +\Pi^{+} (\partial _{+}A^{-}) + \Pi^{-} (\partial_{ +}A^{+} )    
\notag \\
&&  ~~~~~~~~~~~~~~ + \Pi_{u}(\partial_{+}u)  + \Pi_{v}(\partial_{+}v) + \Pi_{w}(\partial_{+}w) - {\cal H}_{T} \biggr ] \notag \\
& = & \biggl[ \frac{1}{2}(\Pi^{-} )^{2} + \partial_{+} \phi^{\dagger} \partial_{-} \phi + \partial_{-} \phi^{\dagger} \partial_{+} \phi + 2 \rho^{2} \phi^{\dagger}\phi A^{+} A^{-} 
\notag \\
&& ~~  - i \rho A^{-} (\phi^{\dagger} \partial_{-}\phi - \phi\partial_{-}\phi^{\dagger}) 
  - i \rho A^{+} (\phi^{\dagger} \partial_{+}\phi - \phi\partial_{+}\phi^{\dagger})  
	- \mu^{2}\phi^{\dagger}\phi - \frac{\lambda}{6} (\phi^{\dagger}\phi)^{2} \biggr] ~~
\label{29}
\end{eqnarray}
In the path integral formulation \cite{29}-\cite{31}, the transition to quantum theory is made  by writing the vacuum to vacuum transition amplitude for the theory called the generating functional $Z[J_{k}]$. For the present theory \cite{25} under the GFC's: $ \zeta_{1} = A^{+} \approx 0 $ and $ \zeta_{2} = A^{-}  \approx 0 $ and in the  presence of the external sources $J_{k}$ it reads:
\begin{eqnarray}
Z[J_{k}] &=& \int [d\mu] \exp \biggl [i \int d^{2} x \biggl [ J_{k} \Phi^{k} + \pi \partial_{+}\phi + \pi^{\dagger} \partial_{+}\phi^{\dagger} + \Pi^{+} \partial _{+}A^{-} \notag \\
&& ~~~~ ~~~~ ~~~~ ~~~~ + \Pi^{-} \partial_{+}A ^{+}    + \Pi_{u}\partial_{+}u  + \Pi_{v}\partial_{+}v + \Pi_{w}\partial_{+} w - {\cal H}_{T} \biggr ] \biggr ] 
\label{30}
\end{eqnarray}
Here, the phase space variables of the theory are: $\Phi^{k} \equiv (\phi, \phi^{\dagger}, A^{-} , A^{+} , u, v, w)$ with the corresponding respective canonical conjugate momenta: $\Pi_{k} \equiv (\pi, \pi^{\dagger}, \Pi^{+} , \Pi^{-} , \Pi_{u},$ $\Pi_{v},\Pi_{w} ) $. The functional measure $[d\mu]$ of the generating functional $Z[ J_{k}]$ under the above gauge-fixing is obtained as : 
\begin{eqnarray}
[d\mu] &=&  [\partial_{-} \delta (x^{-} - y^{-}) ~ \delta(x^{-} - y^{-})][d\phi][d{\phi}^{\dagger}][dA^{+}] [dA^{-}][du][dv][dw] \notag \\      
&& ~~~~ [d\pi][d\pi^{\dagger}][d\Pi^{-}] 
[d\Pi^{+}][d\Pi_{u}][d\Pi_{v}][d\Pi_{w}] 
\delta[\Pi^{+} \approx 0]\delta[ A^{-} \approx 0] \notag \\
&& ~~~~ \delta[(\partial_{-} \Pi^{-} + i \rho (\phi \pi - \phi^{\dagger} \pi^{\dagger}))\approx 0]
\delta [ A^{+} \approx 0] 
\label{31}
\end{eqnarray}
The LF Hamiltonian and path integral quantization of the theory under the set of GFC's: $A^{+} \approx 0 $ and $ A^{-} \approx 0 $ is now complete. 

\section{Spontaneous Symmetry Breaking} 

In this section, we consider the spontaneous symmetry breaking (SSB) in the theory in (i) the so-called unitary gauge and (ii) in the 't Hooft gauge and show explicitly the existence of SSB \cite{32}-\cite{34} in the theory in both the cases.

Our Higgs potential possesses a local maximum at
\begin{equation} 
\phi(x) ~ = ~ \phi_{0} ~ = ~ \sqrt{\biggl(\frac{-3\mu^{2}}{\lambda}\biggr)} ~~~ e^{i\theta} ~~~,~~~ 0 ~ \le ~ \theta < 2\pi 
\label{32}
\end{equation}
where the phase angle $\theta$ defines  a direction in the complex $\phi-$ plane. Here the vacuum state (or the ground state) of the system is clearly non unique, and the SSB will occur for any particular choice of the value of $\theta$. In our considerations we however, choose 
$\theta = 0$ which in turn implies: 
\begin{equation}
\phi_{0} ~ = ~ \sqrt{\biggl(\frac{-3\mu^{2}}{\lambda}\biggr)} 
~ = ~ \frac{v}{\sqrt{2}} 
\label{33}
\end{equation} 
where we choose $v~ > ~ 0$. We now parametrize the field $\phi(x)$ in terms of its deviations from its vacuum expectation value (VEV): $ <0|~ \phi(x) ~|0> = ~ \phi_{0} ~ = ~
(v/\sqrt{2}) ~ > ~ 0 $ in terms of two real fields 
$\sigma(x)$and $\eta(x)$, which measure the deviations of the field $\phi(x)$ from the  equilibrium ground state configuration $\phi(x) = \phi_{0}$. For this we 
expand our complex scalar field $\phi(x)$, in terms of two real fields $\sigma(x)$  and $\eta (x)$ as: 
\begin{equation} 
\phi(x) = [\varphi_{0} + \varphi(x)]
= \frac{1}{\sqrt{2}}[(v + \sigma(x)) + i\eta (x)]
\label{34}
\end{equation}
with
\begin{equation}
\varphi_{0} = \frac{v}{\sqrt{2}} ~~,~~ \varphi(x) = \frac{1}{\sqrt{2}}[(\sigma(x)) + i\eta (x)] 
\label{35}
\end{equation}
such that the real fields $\sigma(x)$ and $\eta(x)$  have vanishing vacuum expectation values. In fact, the term $\varphi_{0}$ here could be interpreted as the zero mode of the theory \cite{32}-\cite{34} and the fluctuation field $\varphi(x)$ could be interpreted as the normal mode of the theory \cite{32}-\cite{34}. 

The Lagrangian density of our LF theory in terms of the real fields $\sigma(x)$ and $\eta(x)$ (after dropping the terms which are irrelevant for our discussions namely, a constant term, a term linear in the field $\sigma(x)$ and all the quartic interaction terms in the fields) (with $~ m_{\sigma} ~ = ~ \sqrt{(2\mu^{2} + \lambda v^{2})/2} ~$ and $~ m_{v} ~ = ~|v\rho| ~$) becomes:
\begin{eqnarray}
{\cal L} &=& \biggl[\frac{1}{2} (\partial_{+}A^{+} - \partial_{-} A^{-})^{2} + 2\partial_{+}\sigma \partial_{-}\sigma + 2\partial_{+} \eta \partial_{-}\eta 
- \frac{1}{2} m_{\sigma}^{2} \sigma^{2} 
\notag \\
&& ~~ + v \rho [A^{+} (x) \partial_{+} \eta(x) 
+ A^{-} (x) \partial_{-}\eta(x)]
- \frac{1}{12} (6\mu^{2} + \lambda v^{2}) \eta^{2}
+ m_{v}^{2} A^{+} A^{-} 
\notag \\
&& ~~ + \rho\sigma[A^{+} (x) \partial_{+} 
\eta(x) + A^{-} (x) \partial_{-}\eta(x)]
- \rho \eta [A^{+} (x) \partial_{+} \sigma(x) 
- A^{-} (x) \partial_{-}\sigma(x)] ~~~~
\notag \\
&& ~~ + 2 v \rho^{2} \sigma A^{+} A^{-}  
- \frac{1}{6}  \lambda v \sigma (\sigma^{2} + \eta^{2}) \biggr] 
\label{36}
\end{eqnarray}
The first term in the above Lagrangian density represents the kinetic energy of the electromagnetic field;  the second term represents the kinetic energy of the real scalar field $\sigma(x)$; the third term represents the kinetic energy of the real scalar field $\eta(x)$; the fourth term represents the mass term for the real scalar field $\sigma(x)$; the fifth term, which involves  the product of the fields $A^{+} (x)$ and $A^{-} (x)$ with the  derivatives of the field $\eta(x)$, represents a quadratic interaction term involving the fields $A^{+} (x), A^{-} (x)$ and $\eta(x)$, and it implies that the fields $A^{+} (x), A^{-} (x)$ and $\eta(x)$ are not independent normal coordinates and are therefore not free fields; consequently the sixth and seventh terms cannot be interpreted as the mass terms for the real scalar field $\eta(x)$ and the electromagnetic field respectively. It also implies that the above Lagrangian density contains an unphysical field; we will eliminate it under some suitable gauge.   The last four terms in the above Lagrangian density represent simply the cubic interaction terms of the theory which will be needed for our later discussions.  

\subsection{The Unitary Gauge and SSB}

We now consider this theory in the so-called unitary gauge.
In fact, for any complex field $\phi(x)$, a gauge transformation can be found which transforms $\phi(x)$ into  a real field such as: 
\begin{equation}
\phi(x) = \frac{1}{\sqrt{2}} [v + \sigma(x)]
\label{37}
\end{equation}
The gauge in which the transformed field has this form is called as the unitary gauge. With this substitution the transformed Lagrangian density in the so-called unitary gauge (after dropping as before, the terms which are irrelevant for our discussions, namely, a constant term,  a term linear in the field $\sigma(x)$ and the quartic interaction terms) becomes:  
\begin{eqnarray}
{\cal L}_{U} &=& {\cal L}_{U}^{0} + {\cal L}_{U}^{int}
\notag \\
{\cal L}_{U}^{0} &=& \biggl[\frac{1}{2} (\partial_{+}A^{+}  - \partial_{-} A^{-} )^{2} + 2\partial_{+}\sigma \partial_{-}\sigma - \frac{1}{2} m_{\sigma}^{2} {\sigma}^{2} 
 + m_{v}^{2} A^{+}  A^{-} \biggr]
\notag \\
{\cal L}_{U}^{int} &=& \biggl[2v \rho^{2} A^{+} A^{-}  \sigma - \frac{\lambda}{6} v \sigma^{3} \biggr] 
\label{38}
\end{eqnarray}
The interaction part of the above Lagrangian density however, does not contain any quadratic coupling terms involving the coupling of the fields $\sigma(x)$, $A^{+} (x)$ and 
$A^{-} (x)$. Hence treating the interaction part of the Lagrangian density ${\cal L}_{U}^{int}(x)$, in perturbation theory, one could interpret ${\cal L}_{U}^{0}$ as the free-field Lagrangian density of a real Klein-Gordon field 
$\sigma(x)$ and a real massive vector field $A_{\mu} (x)$.  Upon quantizing the theory, the field $\sigma(x)$ gives rise to  neutral scalar bosons of mass 
$m_{\sigma} = \sqrt{(2\mu^{2} + \lambda v^{2})/2}$
and the field $A_{\mu} (x)$ gives rise to neutral vector bosons of mass $m_{v} = |v \rho|$. This is an explicit demonstration of the SSB in the theory through the Higgs mechanism where the massive spin $0$ boson associated with the field $\sigma(x)$ is the Higgs boson (or Higgs Scalar) of the theory. Here the vector field $A_{\mu} (x)$ has become massive in the process of SSB through the Higgs mechanism. 

\subsection{The Light-Front 't Hooft Gauge and SSB} 

We consider the LF 't Hooft gauge defined by:
\begin{equation}
[ \partial_{+}A^{+}  + \partial_{-}A^{-} 
- \rho v \eta(x) ] \approx 0 
\label{39}
\end{equation}
and construct the LF 't Hooft-gauge-fixed Lagrangian density of the theory $\tilde{\cal L}$:
\begin{equation}
\tilde{\cal L} = [ {\cal L} + {\cal L}_{'tH} ]
\label{40}
\end{equation}
by adding the LF 't Hooft-gauge- fixing term 
${\cal L}_{'tH}$:
\begin{equation} 
{\cal L}_{'tH} = \biggl[ - \frac{1}{2} [\partial_{+}A^{+}  + \partial_{-}A^{-}  - \rho v \eta(x)]^{2} \biggr]
\label{41} 
\end{equation}
to the Lagrangian density of the theory expressed in terms of the real scalar fields $\sigma$ and $\eta$ 
given by Eq.(33) and ignoring the terms irrelevant for our discussion as explained in the forgoing. The LF 't Hooft gauge-fixed action of the theory $\tilde{S}$ could now be written after a partial integration (and with $m_{\eta} ~ = ~ \sqrt{(6\mu^{2} + \lambda v^{2})/6} ~$~) as:
\begin{eqnarray}
\tilde{S} &=& \int_{ }^{}  \tilde{\cal L} ~ dx^{+} ~ dx^{-}  \notag \\
\tilde{\cal L} &=& \biggl[\frac{1}{2} (\partial_{+}A^{+} - \partial_{-} A^{-} )^{2} + 2\partial_{+}\sigma 
\partial_{-}\sigma + 2\partial_{+} \eta \partial_{-}\eta 
- \frac{1}{2} m_{\sigma}^{2} \sigma^{2} - \frac{1}{2} m_{\eta}^{2} \eta^{2} + m_{v}^{2} A^{+}  A^{-}  \biggr] ~~~
\label{42}
\end{eqnarray}
The fields $\sigma (x)$, $\eta (x)$ and $A^{\mu}(x)$,
could now be treated in perturbation theory as three independent fields which could be quantized in the usual manner.  The LF 't Hooft gauge here reintroduces the field $\eta (x)$ which gets eliminated in the so-called unitary gauge. However, there are no real particles corresponding to the quantized $\eta (x)$ field and they appear in a manner  akin to the longitudinal and scalar photons of QED theory.

\section{Light-Front BRST Quantization} 

For the BRST formulation of the model, we rewrite the theory as a quantum system that possesses the generalized gauge invariance called BRST symmetry.  For this, we first enlarge the Hilbert space of our gauge-invariant theory and replace the notion of gauge-transformation, which shifts operators by c-number functions, by a BRST transformation, which mixes operators with Bose and Fermi statistics. We then introduce new anti-commuting variable c and $ \bar{c} $ 
(Grassman numbers on the classical level and operators in the quantized theory) and a commuting variable $b$ such that \cite{37}-\cite{39}:
\alpheqn
\begin{eqnarray}
\hat{\delta} \phi &=& - i \rho c \phi ~~,~~ \hat{\delta} \phi^{\dagger} = i \rho c \phi^{\dagger} ~~,~~ \hat{\delta} A^{-} = \partial_{+} c ~~,~~ \hat{\delta} A^{+} = \partial_{-} c  \\
\hat{\delta} \pi &=& [{\rho}^{2} c \phi^{\dagger} A^{+}
+ i\rho c \partial_{-}\phi^{\dagger}] 
  ~~,~~ \hat{\delta} \pi^{\dagger} = [{\rho}^{2} c \phi A^{+} - i \rho c \partial_{-}\phi] \\
\hat{\delta} u &=& \hat{\delta} v = \hat{\delta} w = \hat{\delta} \Pi^{+} = \hat{\delta} \Pi^{-} =
\hat{\delta} \Pi_{u} = \hat{\delta} \Pi_{v} = \hat{\delta} \Pi_{w} = 0 \\
\hat{\delta}c &=& 0 ~~,~~ \hat{\delta}\bar{c} = b ~~,~~ 
\hat{\delta}b = 0 
\end{eqnarray}
\label{43}
\reseteqn
with the property $\hat{\delta}^{2}$ = 0.  We now define a BRST-invariant function of the dynamical phase space variables of the theory to be a function $ f $ such that $\hat{\delta} f =  0 $. Now the BRST gauge-fixed quantum Lagrangian density ${\cal L}_{BRST}$ for the theory could be obtained by adding to the first-order Lagrangian density ${\cal L}_{I0}$, a trivial BRST-invariant function, e.g. as follows: 
\begin{eqnarray}
{\cal L}_{BRST} &=&  \biggl[ \frac{1}{2}(\Pi^{-})^{2} + \partial_{+} \phi^{\dagger} \partial_{-} \phi + \partial_{-} \phi^{\dagger} \partial_{+} \phi  
 - i \rho A^{-} (\phi^{\dagger} \partial_{-}\phi - \phi\partial_{-}\phi^{\dagger}) 
- \frac{\lambda}{6} (\phi^{\dagger}\phi)^{2}
\notag \\
&&  - \mu^{2} \phi^{\dagger}\phi
+  2 \rho^{2} \phi^{\dagger}\phi A^{+} A^{-}  - i \rho A^{+} (\phi^{\dagger} \partial_{+}\phi - \phi\partial_{+}\phi^{\dagger}) + \hat{\delta}[\bar{c}(\partial_{ +}A^{-} + \frac{1}{2}b)]\biggr] 
\end{eqnarray}
\label{44}
The last term in the above equation is the extra BRST-invariant gauge-fixing term.  After one integration by parts, the above equation could now be written as:
\begin{eqnarray}
{\cal L}_{BRST} &=&  \biggl[ \frac{1}{2}(\Pi^{-})^{2} + \partial_{+} \phi^{\dagger} \partial_{-} \phi + \partial_{-} \phi^{\dagger} \partial_{+} \phi  
 - i \rho A^{-} (\phi^{\dagger} \partial_{-}\phi - \phi\partial_{-}\phi^{\dagger}) - \mu^{2} \phi^{\dagger}\phi - \frac{\lambda}{6} (\phi^{\dagger}\phi)^{2}
\notag \\
&& ~~ +  2 \rho^{2} \phi^{\dagger}\phi A^{+} 
A^{-}  - i \rho A^{+} (\phi^{\dagger} \partial_{+}\phi - \phi\partial_{+}\phi^{\dagger}) + \partial_{+}A^{-} + \frac{1}{2}b^{2} + (\partial_{+}\bar{c}) (\partial_{+}c) \biggr]
\end{eqnarray}
\label{45}
Proceeding classically, the Euler-Lagrange equation for $b$ reads:
\begin{equation}
- b =(\partial_{ +}A^{-}) 
\end{equation}
\label{46}
the requirement $\hat{\delta} b = 0 $ then implies
\begin{equation}
- \hat{\delta}b =[\hat{\delta}(\partial_{+}A^{-})]
\end{equation}
\label{47}
which in turn implies
\begin{equation}
\partial_{ +}\partial_{ +}c = 0 
\end{equation}
\label{48}
The above equation is also an Euler-Lagrange equation obtained by the variation of ${\cal L}_{BRST}$ with respect to $\bar{c}$. In introducing momenta one has to be careful in defining those for the fermionic variables. We thus define the bosonic momenta in the usual manner so that
\begin{equation}
\Pi^{ +}:= \frac{\partial}{\partial(\partial_{+}A^{-})}
{\cal L}_{BRST}=  b    
\end{equation}
\label{49}
but for the fermionic momenta with directional derivatives we set
\begin{equation}
\Pi_{ c}:= {\cal L}_{BRST}
\frac{\overleftarrow{\partial}}{\partial(\partial_{+}c)} =
\partial _{+}\bar{c} ~~~,~~  \quad \Pi_{\bar{c}} :=
\frac{\overrightarrow{\partial}}{\partial(\partial_{+}\bar{c})} {\cal
L}_{BRST} = \partial_{+}c
\end{equation}
\label{50}
implying that the variable canonically conjugate to $c$ is ( $\partial_{+} \bar{c}$) and the variable conjugate to $\bar{c}$ is ($\partial_{+} c$).  For writing the Hamiltonian density from the Lagrangian density in the usual manner we remember that the former has to be Hermitian so that:
\begin{eqnarray}
{\cal H}_{BRST} &=& \biggl[\pi \partial_{+}\phi +  \pi^{\dagger} \partial_{+}\phi^{\dagger} + \Pi ^{+} \partial_{+}A^{-} +\Pi^{-} \partial_{+}A^{+} + \Pi_{b}\partial_{+}b + \Pi_{u}\partial_{+}u  
\notag \\ 
&& ~~~~~~~~~~~~~~~  + \Pi_{v}\partial _{+}v + \Pi_{w}\partial _{+}w + \Pi_{c}(\partial_{+}c) +(\partial _{+}\bar{c})\Pi_{\bar{ c}} - {\cal L}_{BRST} \biggr]\notag \\
&=& \biggl[ \frac{1}{2}(\Pi^{-})^{2} + \Pi^{-}(\partial_{-}A^{-} - 2 \rho^{2} \phi^{\dagger} \phi A^{+} A^{-} + \mu^{2} \phi^{\dagger}\phi + \frac{\lambda}{6} (\phi^{\dagger}\phi)^{2}
\notag \\
&& ~~~~~~~~~~~~~~~~~~~~~ - i \rho A^{-} (\phi \partial_{-} \phi^{\dagger} - \phi^{\dagger} \partial_{-} \phi) - \frac{1}{2} (\Pi^{+})^{2} + \Pi_{c}\Pi_{\bar{c}} \biggr] 
\end{eqnarray}
\label{51}
The consistency of the last two equations could now be easily checked by looking at the Hamilton's equations for the fermionic variables. Also for the operators $c, \bar{c}, \partial_{+}c$ and $\partial_{+} \bar{c}$, one needs to satisfy the anticommutation relations of $\partial_{+} c$ with $\bar{c}$ or of $\partial_{+} \bar{c}$ with $c$, but not of $c$, with $\bar{c}$. In general, $c$ and $\bar{c}$ are independent canonical variables and one assumes that \cite{37}-\cite{39}:
\begin{equation}
\{ \Pi_{c}, \Pi_{\bar{c}} \} = \{ \bar{c},c \} = \partial_{+} \{ \bar{c}, c \} = 0  ~,~
\{ \partial_{+} \bar{c} , c \}  =  (-1) \{ \partial_{+}c, \bar{c} \}
\end{equation}
\label{52}
where $\{\makebox{ }$,  $\}$ means an anticommutator.  We thus see that the anticommulators in the above equation are non-trivial and need to be fixed.  In order to fix these, we demand that c satisfy
the Heisenberg equation:
\begin{equation}
[c, {\cal H}_{BRST}] = i \partial_{+} c 
\end{equation}
\label{53}
and using the property $c^{2}$ = $ \overline{c}^{2}$ = 0 one obtains
\begin{equation}
[c, {\cal H}_{BRST}]= \{ \partial_{+} \bar{c}, c \}
\partial_{+}c                
\end{equation}
\label{54}
The last three equations then imply :
\begin{equation}
\{ \partial_{ +} \bar{c} , c \} = (-1) \{ \partial_{ +} c, \bar{c} \} = i 
\end{equation}
\label{55}
Here the minus sign in the above equation is nontrivial and implies the existence of states with negative norm in the space of state vectors of the theory. The BRST charge operator $Q$ is the generator of the BRST transformations. It is nilpotent and satisfies $Q^{2}$ = 0. It mixes operators which satisfy Bose and Fermi statistics. According to its conventional definition, its commutators with Bose operators and its anti-commutators with Fermi operators for the present theory satisfy:
\alpheqn 
\begin{eqnarray}
[\phi , Q] &=& - i \rho \phi c   ~,~
 [\phi^{\dagger} , Q] = i \rho {\phi}^{\dagger} c \\
\label{56a}
[\pi , Q] &=& - i \rho c \pi ~,~
 [\Pi^{\dagger} , Q] = i \rho c {\pi}^{\dagger}  \\
\label{56b} 
[A^{+}, Q] &=& \partial_{-} c ~,~ [A^{-}, Q] = \partial_{+} c 
~,~ [\Pi^{+} , Q] = [\Pi^{-} , Q] = 0 \\
\label{56c} 
\{\partial_{+}\bar{c},Q\} &=& [-\partial_{-} \Pi^{-} - i \rho (\phi\pi - \phi^{\dagger}\pi^{\dagger}) ] 
~,~ \{ \bar{c},Q\} =  (- \Pi^{+})
\label{56d}
\end{eqnarray}
\reseteqn
All other commutators and anti-commutators involving $Q$ vanish. In view of this, the BRST charge operator of the present  theory can be written as: 
\begin{equation}
Q = \int_{ }^{}\!dx^{-}
\biggl[ i c ~\partial_{-} \Pi^{-} - \rho c (\phi\pi - \phi^{\dagger} \pi^{\dagger}) 
- i \partial_{+} c ~\Pi^{+} \biggr]
\end{equation} 
\label{57}
This equation implies that the set of states satisfying the conditions: 
\begin{equation}
\Pi^{+} |\psi \rangle = 0 ~~,~~ 
[\partial_{-} \Pi^{-} + i \rho (\phi \pi
- \phi^{\dagger} \pi^{\dagger})] |\psi \rangle = 0
\end{equation}
\label{58}
belong to the dynamically stable subspace of states $|\psi>$ satisfying $Q|\psi> = 0 $, i.e., it belongs to the set of BRST-invariant states. In order to understand the condition needed for recovering the physical states of the theory we rewrite the operators $c$ and $\bar{c}$ in terms of fermionic annihiliation and creation operators. For this purpose we consider Euler lagrange equation for the variable $ c$ derived earlier. The solution of this equation gives (for the light-cone time $x^{+} \equiv \tau )$  the Heisenberg operators $c(\tau)$ and correspondingly  $\bar{c}(\tau)$ in terms of the fermionic annihilation and creation operators as:
\begin{equation}
c(\tau) = G(\tau) + F(\tau), ~~~ \bar{c}(\tau) = G^{\dagger}(\tau) + F^{\dagger}(\tau)
\end{equation}
\label{59}
Which at the light-cone time $\tau = 0$ imply
\alpheqn
\begin{eqnarray}
c &\equiv& c(0) = F, \quad \bar{c}(\tau) \equiv \bar{c}(0) = F^{\dagger}\\
\partial_{+} c(\tau) &\equiv& \partial_{+} c(0) = G, \quad \partial_{+} \bar{c}(\tau) \equiv \partial_{+} \bar{c}(0) = G^{\dagger}
\end{eqnarray}
\label{60}
\reseteqn
By imposing the conditions (obtained earlier):
\alpheqn
\begin{eqnarray}
c^{2}  = \bar{c}^{2} = \{ \bar{c} , c \} = \{ \partial_{+} \bar{c}, \partial_{+}c \} &=& 0  \\
\{ \partial_{ +} \bar{c} , c  \} = (-1) \{ \partial_{+}c , \bar{c} \} &=& i 
\end{eqnarray}
\label{61}
\reseteqn
we then obtain
\begin{equation}
F^{2}  =  (F^{\dagger})^{2} = \{ F^{\dagger}, F \} = \{ G^{\dagger}, G \} = 0 ~,~  \{ G^{\dagger}, F \} = (-1) \{ G, F^{\dagger} \} = i
\label{62}
\end{equation}

Now let $|0>$ denote the fermionic vacuum for which
\begin{equation}
G|0> = F|0> = 0
\end{equation}
\label{63}
Defining $|0>$ to have norm one, the last three equations imply
\begin{equation}
<0|F {G^{\dagger}}|0> = i   ~~~~,~~~~ <0|G {F^{\dagger}}|0> = - i 
\end{equation}
\label{64}
so that 
\begin{equation}
G^{\dagger} |0> \ne 0 ~~~~,~~~~ F^{\dagger} |0> \ne 0
\end{equation}
\label{65}
The theory is thus seen to possess negative norm states in the fermionic sector.  The existence of these negative norm states as free states of the fermionic part of $\cal {H}_{BRST}$ is however, irrelevant to the existence of physicsl states in the orthogonal subspace of the Hilbert space. In terms of annihilation and creation operators ${\cal {H}}_{BRST} $ is:
\begin{eqnarray}
{\cal H}_{BRST} &=& \biggl[ \frac{1}{2}(\Pi^{-})^{2} 
+ \Pi^{-}(\partial_{-}A^{-}) 
+ \mu^{2} \phi^{\dagger}\phi + \frac{\lambda}{6} (\phi^{\dagger}\phi)^{2} - 2 \rho^{2} \phi^{\dagger} 
\phi A^{+} A^{-} \notag \\
&& ~~~~ ~~~~~ ~~~~ ~~~~~~~ ~~~~ - i \rho A^{-} (\phi \partial_{-} \phi^{\dagger} - \phi^{\dagger} \partial_{-} \phi) - \frac{1}{2} (\Pi^{+})^{2} 
+ G^{\dagger}G \biggr] 
\end{eqnarray}
\label{66}
and the BRST charge operator is: 
\begin{equation}
Q = \int_{ }^{}\!dx^{-}
\biggl[ i F ~ \partial_{-} \Pi^{-} - \rho F (\phi\pi - \phi^{\dagger} \pi^{\dagger}) 
- i G ~\Pi^{+} \biggr]
\end{equation} 
\label{67}
Now because $ Q|\psi> = 0 $, the set of states annihiliated by $Q$ contains not only the set for which the constraints of the theory hold but also additional states for which 
\begin{equation}
F|\psi>  =  G |\psi> = 0 ~~,~~ 
\Pi^{+} |\psi \rangle \ne 0   ~~,~~  
[\partial_{-} \Pi^{-} + i \rho (\phi \pi
- \phi^{\dagger} \pi^{\dagger})] |\psi \rangle \ne 0 
\end{equation}
\label{68}
The Hamiltonian is also invariant under the anti-BRST transformation given by:
\alpheqn
\begin{eqnarray}
\bar{\hat{\delta}} \phi &=& i \rho \bar{c} \phi ~~,~~ \bar{\hat{\delta}} \phi^{\dagger} = - i \rho \bar{c} \phi^{\dagger} ~~,~~ \bar{\hat{\delta}} A^{-} = - \partial_{+} \bar{c} ~~,~~ \bar{\hat{\delta}} A^{+} = - \partial_{-} \bar{c} \\
\bar{\hat{\delta}} \pi &=& [ -{\rho}^{2} \bar{c} \phi^{\dagger} A^{+}
- i\rho \bar{c} \partial_{-}\phi^{\dagger}] 
  ~~,~~ \bar{\hat{\delta}} \pi^{\dagger} = [ - {\rho}^{2} \bar{c} \phi A^{+} + i \rho \bar{c} \partial_{-}\phi] \\
\bar{\hat{\delta}} u &=& \bar{\hat{\delta}} v = \bar{\hat{\delta}} w = \bar{\hat{\delta}} \Pi^{+} = \bar{\hat{\delta}} \Pi^{-} =
\bar{\hat{\delta}} \Pi_{u} = \bar{\hat{\delta}} \Pi_{v} = \bar{\hat{\delta}} \Pi_{w} = 0 \\
\bar{\hat{\delta}}c &=& - b ~,~ \bar{\hat{\delta}}\bar{c} = 0 ~,~ \bar{\hat{\delta}}b = 0 
\end{eqnarray}
\label{69}
\reseteqn
with generator or anti-BRST charge
\begin{equation}
\bar{Q} = \int_{ }^{}\!dx^{-}
\biggl[ - i \bar{c}~ \partial_{-} \Pi^{-} - \rho \bar{c} (\phi\pi - \phi^{\dagger} \pi^{\dagger}) 
+ i \partial_{+} \bar{c} ~\Pi^{+} \biggr]
\end{equation}
\label{70}
which in terms of annihilation and creation operators reads: 
\begin{equation}
\bar{Q} = \int_{ }^{}\!dx^{-}
\biggl[ - i F^{\dagger} ~\partial_{-} \Pi^{-} - \rho F^{\dagger} (\phi\pi - \phi^{\dagger} \pi^{\dagger}) + i G^{\dagger} ~\Pi^{+} \biggr]
\end{equation} 
\label{71}
We also have 
\begin{equation}
\partial_{ +}Q= [Q, H_{BRST}]= 0 ~~~,~~~
\partial_{ +}\bar{Q}=
[\bar{Q}, H_{BRST}]= 0 
\end{equation}
\label{72}
with
\begin{equation}
H_{BRST}= \int_{ }^{} dx^{-} {\cal H}_{BRST}
\end{equation}
\label{73}
and we further impose the dual condition that both $Q$ and
$\bar{Q}$ annihilate physical states, implying that:
\begin{equation}
Q |\psi > =0  ~~~ {\rm and} ~~~   \bar{Q}|\psi> = 0 
\end{equation}
\label{74}
The states for which the constraints of the theory hold, satisfy both of these conditions and are in fact, the only states satisfying both of these conditions, since although with 
\begin{equation}
 G^{\dagger} G  =  (-1) G G^{\dagger}
\end{equation}
\label{75}
there are no states of this operator with $G^{\dagger}|\psi>  = 0 $ and $F^{\dagger}|\psi> = 0$, and hence no free eigenstates of the fermionic part of $\cal{H}_{BRST} $ that are annihilated by each of $G$, $G^{\dagger}$,
$F$, and  $ F^{\dagger}$. Thus the only states satisfying  $Q|\psi> = 0 $ and $ \bar{Q}|\psi> = 0 $ are those that satisfy the constraints of the theory. 
 
Now because $Q|\psi > = 0$, the set of states annihilated by $Q$ contains not only the set of states for which the constraints of the theory hold but also additional states for which the constraints of the theory do not hold. This situation is, however, easily avoided by additionally imposing on the theory, the dual condition: $Q|\psi> = 0$ and 
$\bar{Q}|\psi> = 0$. By imposing both of these conditions  on the theory simultaneously, one finds that the states for which the constraints of the theory hold are the only states satisfying both of these conditions. This is traced to the conditions on the fermionic variables $c$ and $\bar{c}$ which constrain the solutions such that one cannot have simultaneously $c$, $\partial_{+} c$ and $\bar{c}$, 
$\partial_{+}\bar{c}$, applied to $|\psi >$ giving zero. 
Thus the only states satisfying $Q |\psi> = 0$ and 
$\bar{Q}|\psi> = 0$ are those that satisfy the constraints of the theory and they belong to the set of BRST-invariant as well as to the set of anti-BRST-invariant states.

Alternatively, one can understand the above point in terms of fermionic annihiliation and creation operators as follows.  The condition  $Q|\psi > = 0$ implies the that the set of states annihilated by $Q$ contains not only the states for which the constraints of the theory hold but also additional states for which the constraints do not hold. However, $\bar{Q}|\psi> = 0$ guarantees that the set of states annihilated by  
$\bar{Q}$ contains only the states for which  the constraints  hold, simply because $ G^{\dagger}|\psi> \ne 0$ and 
$F^{\dagger}|\psi>  \ne 0$. This completes the BRST formulation of the theory.

\section{Summary and Discussion}

Theoretical and experimental studies of multi-quark states are challenging and a number of phenomenological models \cite{1}-\cite{26} have been proposed in order to provide interpretation and gain understanding.  

Some of the states \cite{2, 3}, \cite{9}-\cite{19} which do not fit in to the standard classification of mesons (two quark states) and baryons (three quark states) \cite{1}-\cite{3} 
find a rather more natural interpretation in terms of the tetra-quark states or the diquark-antidiquark states \cite{2, 3}, \cite{9}-\cite{25}. 

In particular, as mentioned in the foregoing, Grinstein, Jora and Polosa \cite{25} have studied a model of large-$N$  scalar $QCD_{2}$ \cite{25}. This theory of Grinstein et al. \cite{25} admits a Bethe-Salpeter equation describing the discrete spectrum of $q\bar{q}$ bound states \cite{20}-\cite{25}. In the their work, the gauge fields have been considered \cite{25} in the adjoint representation of $SU(N)$ and the scalar fields in the fundamental representation. The theory is asymptotically free and linearly confining \cite{25}. Different aspects of this theory have been studied by several authors in various contexts \cite{20}-\cite{25}.

In Ref.\cite{26}, we have studied the LFQ of the theory of large-$N$ scalar $QCD_{2}$ studied by Grinstein, Jora and Polosa \cite{25}, \textit{without} Higgs potential\cite{20}-\cite{25} on the LF using the Hamiltonian \cite{28} and path integral \cite{29}-\cite{31} formulations. In in the present work, we have studied this theory \textit{in the presence of} a Higgs potential and we have studied its LFQ using the Hamiltonian, path integral and BRST formulations
\cite{37}-\cite{39}. We have also  shown  explicitly  the occurrence of the SSB in the theory in the unitary gauge as well as in the LF 't Hooft gauge \cite{32}-\cite{34}. 

In the Hamiltonian and path integral quantization of the theory the gauge-invariance of the theory gets broken because
of the gauge-fixing. In view of this, we go to a more generalized quantization procedure called the BRST quantization \cite{37}-\cite{39}, where the extended gauge symmetry of the theory is maintained even under gauge-fixing. 

In the present work, we have studied the LF-BRST quantization of the theory under some specific LF-BRST gauge-fixing (where a particular but non-unique gauge has been chosen). In this procedure, we embed the original GI  theory into a BRST system, the quantum Hamiltonian $H_{BRST}$ (which includes the gauge-fixing contribution) commutes with the BRST charge as well as with the anti-BRST charge. The new extended gauge symmetry which replaces the gauge invariance is maintained (even under the BRST gauge-fixing) and projecting any state onto the sector of BRST and anti-BRST invariant states yields a theory which is isomorphic to the original GI theory.

\section{Acknowledgments}

The authors thank Stan Brodsky for the motivations and continuous support and for his collaboration in the early stages of this work and especially for his contributions on the SSB aspects of this work and for several clarifying suggestions and discussions as well as for providing useful references on the subject. This work was supported in part by the US Department of Energy under Grant No. DE-FG02-87ER40371.


\newpage

\begin{thebibliography}{99}

\bibitem{1} M. Gell-Mann, Phys. Lett. {\bf 8}, (1964) 214; E. Witten, Nucl. Phys. {\bf B160}, (1979) 57; for a good review of the subject, see: S. J. Brodsky, H. C. Pauli and S.S. Pinsky, Phys. Rep. {\bf 301}, (1998) 299.

\bibitem{2} R. L. Jaffe, Phys. Rev. {\bf D 15}, 267 (1977); R. L. Jaffe, Phys. Rev. D 15, 281 (1977).

\bibitem{3} R. L. Jaffe, Phys. Rep. {\bf 409}, (2005) 1; R. L. Jaffe and F. Wilczek, Phys. Rev. Lett. {\bf 91}, (2003) 232003.

\bibitem{4} G. 't Hooft, Nucl. Phys. {\bf B75}, (1974) 461.

\bibitem{5} Chung-I Tan and Zheng Xi-te, ``Gauge-Invariant Treatment of Two-Dimensional Quantum Chromodynamics in the Large-$N$ Limit'', Phys. Rev. {\bf 26D} (No. 10), (1982) 2827.

\bibitem{6} S. J. Brodsky and H. C. Pauli, ``Light-Cone Quantization of Quantum Chromo-dynamics'', Report No. SLAC-PUB-5558, (1991) (unpublished).

\bibitem{7} A. C. Kalloniatis, H. C. Pauli and S. Pinsky, Phys. Rev. {\bf D50}, (1994) 6633. 

\bibitem{8} Hiroyuki Fujita, Sh.M. Shvartsman, ``Role of zero modes in quantization of QCD in light cone coordinates'', Case Western Reserve University Preprint, CWRU-TH-95-1es 1, (22 pp), Jun 1995 (arXiv: hep-th/9506046). 

\bibitem{9} A. D. Polosa, ``Hints of a New Spectroscopy'', {\bf eConf C070805},  (2007) 36.

\bibitem{10} N.V. Drenska, R. Faccini, A.D. Polosa, 
``Higher Tetraquark Particles'', Phys.Lett. {\bf B669}, (2008) 160-166.

\bibitem{11} N.V. Drenska, R. Faccini, A.D. Polosa, 
``Exotic Hadrons with Hidden Charm and Strangeness'',
Phys.Rev. {\bf D79}, (2009) 077502.

\bibitem{12} E.N.M. Cirillo, M. Mori, A.D. Polosa, ``The Delta-Statistics of Unconventional Quarkonium-like Resonances'', Phys. Lett. {\bf B705}, (2011) 498-502.

\bibitem{13} Riccardo Faccini, Alessandro Pilloni, Antonio D. Polosa, ``Exotic Heavy Quarkonium Spectroscopy: A Mini-review'', Mod. Phys. Lett. {\bf A27}, (2012) 1230025.

\bibitem{14} L. Maiani, A.D. Polosa, V. Riquer,
``Indications of a Four-Quark Structure for the X(3872) and 
X(3876) Particles from Recent Belle and BABAR Data'', 
Phys. Rev. Lett. {\bf 99}, (2007) 182003;  
C. Bignamini, B. Grinstein, F. Piccinini, A. Polosa and C. Sabelli, ``Is the $X(3872)$ Production Cross Section at Tevatron Compatible with a Hadron Molecule Interpretation?'', Phys. Rev. Lett. {\bf 103}, (2009) 162001.  

\bibitem{15} T.J. Burns, F. Piccinini, A.D. Polosa, C. Sabelli, ``The $2^{-+}$ Assignment for the X(3872)'', Phys. Rev.  {\bf D82}, (2010) 074003.

\bibitem{16} Federico Brazzi, Benjamin Grinstein, Fulvio Piccinini, Antonio Davide Polosa, Chiara Sabelli, ``Strong Couplings of $X(3872)_{J=1,2}$ and a New Look at 
$J/\psi$ Suppression in Heavy Ion Collisions'', 
Phys. Rev. {\bf D84}, (2011) 014003.

\bibitem{17} R. Faccini, F. Piccinini, A. Pilloni, A.D. Polosa, ``On the Spin of the $X(3872)$'', Phys. Rev. {\bf D86}, (2012) 054012.

\bibitem{18} Stanley J. Brodsky, D. S. Hwang and R. F. Lebed,  
``A New Picture for the Formation and Decay of the Exotic
XYZ Mesons'', Phys. Rev. Lett. {\bf 113}, (2014) 112001
(SLAC-PUB-16001), (hep-ph/1406.7281). 

\bibitem{19} Richard J. Lloyd and James P. Vary, ``All-charm Tetraquarks'', Phys. Rev. {\bf D70} (2004) 014009.



\bibitem{20} L. Maiani, F. Piccinini, A.D. Polosa, V. Riquer,  ``A New look at Scalar Mesons'', Phys. Rev. Lett. {\bf 93}, (2004) 212002. 

\bibitem{21} I. Bigi, L. Maiani, F. Piccinini, A.D. Polosa, V. Riquer, ``Four-quark Mesons in Non-leptonic B Decays: Could they Resolve Some Old Puzzles?'', 
Phys. Rev. {\bf D72}, (2005) 114016.

\bibitem{22} Luciano Maiani, Fulvio Piccinini, Antonio D. Polosa, Veronica Riquer, ``Diquark-antidiquark states with hidden or open charm'', {\bf PoS} HEP2005, (2006) 105.

\bibitem{23} A. D. Polosa, ``Scalar Mesons Dynamics'', Nucl. Phys. Proc. Suppl. {\bf 181-182}, (2008) 175-178.

\bibitem{24} G. 't Hooft, G. Isidori, L. Maiani, A.D. Polosa, V. Riquer, ``A Theory of Scalar Mesons'', Phys.Lett. {\bf B662}, (2008) 424-430.

\bibitem{25} Benjamin Grinstein, R. Jora, A.D. Polosa, 
``A Note on Large N Scalar QCD(2)'', Phys.Lett. 
{\bf B671}, (2009) 440-444. 

\bibitem{26} Usha Kulshreshtha, D. S. Kulshreshtha and J. P. Vary, ``Light-front Hamiltonian and Path Integral Formulations of Large N Scalar $QCD_{2}$'', 
Phys. Lett. {\bf B708}, (2012) 195-198.

\bibitem{27} D. S. Kulshreshtha, ``LFQ of Large N Scalar 
$QCD_{2}$ with a Higgs Potential'', Invited Talk at the International Conference on Nuclear Theory in the Supercomputing Era-2013 (NTSE-2013), held at the Iowa State University, Ames, Iowa, USA, May 13-17 (2013), Published in the Conference Proceedings (NTSE-2013).


\bibitem{28} P. A. M. Dirac, Can. J. Math {\bf 2}, (1950), 129.

\bibitem{29} P. Sanjanovic, Annals Phys. (N.Y.) {\bf 100}, (1976) 227.

\bibitem{30} U. Kulshreshtha and D. S. Kulshreshtha, Phys. Lett. {\bf B555}, (2003) 255. 

\bibitem{31} U. Kulshreshtha and D. S. Kulshreshtha, European Phys. Jour. {\bf C29}, (2003) 453.

\bibitem{32} Prem P. Srivastava, Stanley J. Brodsky,
``Light Front Quantized QCD in Light Cone Gauge'',
Phys. Rev. {\bf D64} (2001) 045006.

\bibitem{33} Prem P. Srivastava, Stanley J. Brodsky,
``Light Front Quantized QCD in Covariant Gauge '',
Phys. Rev. {\bf D61} (2000) 025013. 

\bibitem{34} Prem P. Srivastava and Stanley J. Brodsky,
``A Unitary and Renormalizable Theory of the Standard Model in Ghost Free Light Cone Gauge'',
Phys.Rev. {\bf D66} (2002) 045019.

\bibitem{35} P. A. M.  Dirac, Rev. Mod. Phys.{\bf  21}, 
(1949) 392.

\bibitem{36} S. J. Brodsky, H. C. Pauli and S.S. Pinsky, Phys. Rep. {\bf 301}, (1998) 299.

\bibitem{37} C. Becchi, A. Rouet and A. Stora, Phys. Lett. {\bf B52}, 344 (1974).

\bibitem{38} V. Tyutin, Lebedev {\bf Report No. FIAN-39}, (unpublished).

\bibitem{39} D. Nemeschansky, C. Preitschopf and M. Weinstein, Ann. Phys.(N.Y.) {\bf 183}, 226 (1988). 

\bibitem{40} J. S. Rozowsky and C. B. Thorn, ``Spontaneous Symmetry Breaking at Infinite Momentum without P+ Zero Modes'', 
Phys. Rev. Letts. {\bf 85}, 1614-1617 (2000) (arXiv: hep-th/0003301). 
 
\bibitem{41} D. Chakrabarti, A. Harindranath and J. P. Vary,
``A Transition in the Spectrum of the Topological Sector of 
$\phi_{2}^{4}$ Theory at Strong Coupling'', Phys. Rev. 
{\bf D71}, 125012 (2005) (arXiv: hep-th/0504094).
 
\bibitem{42} D. Chakrabarti, A. Harindranath, L. Martinovic and J. P. Vary, ``Kinks in Discrete Light-Cone Quantization'', Phys. Letts. {\bf B582} (2004) 196-202 (2004).

\bibitem{43} D. Chakrabarti, A. Harindranath, L. Martinovic, G. B. Pivovarov and J. P. Vary, ``Ab initio Results for the Broken Phase of Scalar Light-Front Field Theory'', Phys. Letts. {\bf B617} 92-98 (2005) (arXiv: hep-th/0310290).

\bibitem{44} V.T. Kim, G. B. Pivovarov, J.P. Vary, ``Phase Transition in Light-Front $\phi^{4}$ (1+1)'', Phys.Rev. {\bf D69}, 085008 (2004) (arXiv: hep-th/0310216).

\bibitem{45} D. Chakrabarti, A. Harindranath, L. Martinovic 
and J.P. Vary, ``Kinks in Discrete Light-Cone Quantization'', Phys.Lett. B582 (2004) 196-202 (arXiv: hep-th/0309263).

\bibitem{46} A. Harindranath and J. P. Vary, ``Light-Front Hamiltonian Approach to Relativistic Two and Three-body Bound State Problems in (1+1)-Dimensions '', Phys. Rev. {\bf D37}, 1064-1069 (1988).
 
\bibitem{47} A. Harindranath, J.P. Vary, ``Stability of the Vacuum in Scalar Field Models in 1 + 1 Dimensions'',
Phys.Rev. {\bf D37}, (1988) 1076-1078 (1988).

\bibitem{48} A. Harindranath, J.P. Vary, ``Solving Two-dimensional $\Phi^{4}$ Theory By Discretized Light-Front Quantization'', Phys.Rev. {\bf D36} (1987) 1141-1147 (1987).

\end{thebibliography}
\end{document}